\documentclass[twocolumn,prx,superscriptaddress,notitlepage,longbibliography]{revtex4-1}

\usepackage{graphicx}
\usepackage{rotate}
\usepackage{epsfig}
\usepackage{xcolor}
\usepackage[normalem]{ulem}
\usepackage{amssymb,amsfonts,amsmath}

\usepackage{todonotes}

\newcommand{\tb}{\textcolor{blue}}

\begin{document}

\title{Individual perception dynamics in drunk games}

\author{Alberto Antonioni}
\thanks{authors contributed equally and appear alphabetically}
\affiliation{Department of Economics, University College London, UK}
\affiliation{Grupo Interdisciplinar de Sistemas Complejos (GISC), Departamento de Matem\'aticas, Universidad Carlos III de Madrid, E-28911~Legan\'es, Madrid, Spain}
\affiliation{Institute for Biocomputation and Physics of Complex Systems (BIFI), University of Zaragoza, E-50018 Zaragoza, Spain}
\email{alberto.antonioni@gmail.com}

\author{Luis A. Martinez-Vaquero}
\thanks{authors contributed equally and appear alphabetically}
\affiliation{Institute of Cognitive Sciences and Technologies, National Research Council of Italy (ISTC-CNR), 00185 Rome, Italy}
\email{l.martinez.vaquero@gmail.com}

\author{Cole Mathis}
\thanks{authors contributed equally and appear alphabetically}
\affiliation{Beyond Center for Fundamental Questions in Science,  Arizona State University, Tempe AZ, USA}
\affiliation{Department of Physics,  Arizona State University, Tempe AZ, USA}
\email{cole.mathis@asu.edu}

\author{Leto Peel}
\thanks{authors contributed equally and appear alphabetically}
\affiliation{ICTEAM, Universit\'{e} catholique de Louvain, Avenue George Lema\^{i}tre 4, B-1348 Louvain-la-Neuve, Belgium}
\email{piratepeel@gmail.com}
\affiliation{naXys, Universit\'{e} de Namur, Rempart de la Vierge 8, 5000 Namur, Belgium}

\author{Massimo Stella}
\thanks{authors contributed equally and appear alphabetically}
\affiliation{Institute for Complex Systems Simulation, Southampton, UK}
\email{massimo.stella@inbox.com}

\pacs{}

%
%

\begin{abstract}
We study the effects of individual perceptions of payoffs in two-player games. In particular we consider the setting in which individuals' perceptions of the game are influenced by their previous experiences and outcomes. Accordingly, we introduce a framework based on evolutionary games where individuals have the capacity to perceive their interactions in different ways. Starting from the narrative of social behaviors in a pub as an illustration, we first study the combination of the prisoner's dilemma and harmony game as two alternative perceptions of the same situation. Considering a selection of game pairs, our results show that the interplay between perception dynamics and game payoffs gives rise to non-linear phenomena unexpected in each of the games separately, such as catastrophic phase transitions in the cooperation basin of attraction, Hopf bifurcations and cycles of cooperation and defection. Combining analytical techniques with multi-agent simulations we also show how introducing individual perceptions can cause non-trivial dynamical behaviors to emerge, which cannot be obtained by analyzing the system as a whole. Specifically, initial heterogeneities at the microscopic level can yield a polarization effect that is unpredictable at the macroscopic level. This framework opens the door to the exploration of new ways of understanding the link between the emergence of cooperation and individual preferences and perceptions, with potential applications beyond social interactions.
\end{abstract}

\maketitle

\section{Introduction}

Game theory provides a useful mathematical formalism to investigate the logical decision-making processes of intelligent, rational individuals that maximize their expected payoff in conflicting interest situations~\cite{nash1950equilibrium,vega-redondo-03}. 
In simple non-cooperative games, cooperators can be vulnerable to exploitation by selfish partners and so the dominant rational behavior is expected to be uncooperative~\cite{nowak2006evolutionary,sigmund}, as originally conjectured by Darwin~\cite{darwin:1871}. However, cooperative behavior is observed at practically every level of biological and societal organization~\cite{hammerstein:2003}, playing a key role in the major steps of evolution~\cite{maynard-smith:1995}.
Controlled laboratory and field experiments have also measured non-negligible amounts of cooperative behavior among humans~\cite{andreoni95,fehr2000,milinski2002,Gracia,Grujic,antonioni2016cooperation}. Accordingly, many mechanisms have been proposed to explain the emergence of cooperation in both animal and human societies~\cite{hamilton1964,nowak2006five}. 
In this work we consider the effect of heterogeneous individual perceptions in games on the evolution of cooperation. 

It has been argued that individuals do not necessarily play rationally, but instead rational behaviors may emerge through forms of adaptation. Thus far, two distinct mechanisms for adaptive systems have been proposed: \textit{learning} and \textit{evolution}~\cite{fudenberg1998theory,sutton1998reinforcement,smith1973logic}. Learning focuses on the local optimization of individual strategies, whereas evolution considers the adaptation of whole populations of individuals. In learning systems, individuals ``learn'' their strategies over repeated games by choosing actions to directly maximize their expected payoff~\cite{fudenberg1998theory}. The study and development of optimal learning strategies has become a subject of interest within the field of machine learning, particularly when the payoffs are stochastic or unknown~\cite{sutton1998reinforcement}. In evolutionary game theory (EGT) players have fixed strategies, but asexually reproduce offspring with strategies proportional to their utility~\cite{smith1973logic}. In social and economic settings, where individuals do not reproduce, this mechanism can be interpreted as a form of social learning in which individuals imitate those with higher utilities. 
EGT has been proven to be a powerful tool to study the emergence of cooperation in a broad range of problems in which dilemmas are present~\cite{binmore1994game,friedman1998economic,dugatkin2000game,nowak2004}.
 
Common to many of these theoretical adaptive systems is an inherent assumption of homogeneity that all individuals value the payoffs of specific outcomes in an identical manner. In other words, game payoffs only depend on the set of actions played, are invariant between individuals and remain constant over time. However, there is evidence that suggests individuals perceive equivalent outcome scenarios differently. For instance, experiments on populations from different cultures indicate that individuals appear to assign different values to the prescribed payoffs through an implicit ``mapping'' of the game to social exchanges that are more familiar to them~\cite{henrich2001search,oosterbeek2004cultural}. 
 
Within this work we postulate that individuals may have different perceptions of the same set of outcomes and that these perceptions are shaped by their previous experience of the game. For example, there has been a long-standing view that trust can promote cooperation between organizations and/or individuals~\cite{putnam1994making, kreps1982rational} and trust can be built or broken based on prior interactions. There may be a number of mechanisms that change perceptions such as diminishing returns for repeated actions, e.g., the benefit of scoring points in sports can change depending on whether a team is currently leading or not~\cite{peel2015predicting}. Perceived benefits of competing technologies can vary between individuals and may change over time as a function of those investing in the technology~\cite{arthur1989competing}. We present a framework to model these types of systems by allowing individuals to have different perceptions. Perceptions are modeled as different sets of payoffs and the set of payoffs perceived by an individual is determined by their state, which is dynamically influenced by past experience.

Mixed games, in which individuals in the population play one of two possible games, have previously been considered~\cite{cressman2000evolutionary, hashimoto2006unpredictability, amaral2016evolutionary, wardil2013evolution, wang2014different}. However, in most cases these mixed games transpire to produce the same average behavior as the weighted mean of the two games. Mixed games are different in structured populations~\cite{amaral2016evolutionary}, where the average game is returned only if heterogeneity in payoffs is small. Dynamic payoffs have also been considered within dynamical games, in which payoffs are coupled with the evolution of time~\cite{akiyama2000,akiyama2002,cherkashin2009}, or games in which payoffs are coupled with population strategies~\cite{stewart2014,weitz2016}. However, within all these scenarios, individuals perceive rewards independent of their own specific experiences.

A number of mechanisms have been shown to facilitate cooperation in evolutionary settings, but many of these necessitate an infeasible level of complexity or cognitive load, such as high memory capacity~\cite{hilbe2017memory} or recognition of the others~\cite{taborsky2016correlated}, to occur in natural scenarios~\cite{hammerstein:2003b,WHITLOCK:2007kz, Hauser:2009ju,Andre:2014kda}. Frameworks incorporating states of individual players have been used to reduce the complexity of such mechanisms~\cite{van2012evolution}.  Player states in these models are typically used to directly modify the actions individuals choose. In contrast, our proposed framework uses states to modify the way each individual perceives the game and uses simple strategies that do not have direct dependence on the current state.  

Here we introduce \textit{drunk game theory} (DGT), a framework that couples games to allow individuals to change their \textit{perception} (i.e., the game they play) according to their own prior experience (i.e., the outcomes of their previous games). We first exemplify DGT with a particular scenario called the \textit{Pub Dilemma} that couples a Prisoner's Dilemma and a Harmony game.

In the following, we briefly review some key notions from the study of two-player two-strategy symmetric games~(Sec.~\ref{sec_two_player}). We then introduce the \textit{Pub Dilemma} and provide a generalization for any other pair of games~(Sec.~\ref{sec_drunk_games}). We demonstrate analytically the emergence of new fixed points and critical phase transitions and show that stable fixed points in the original games can lose their stability in the resulting coupled game~(Sec.~\ref{sec_dynamics_drunk_games}). Subsequently, we confirm analytical results in agent-based simulations and extend them to the study of individual behavior~(Sec.~\ref{sec_agent_based}). Finally, we discuss the wide range of potential multidisciplinary applications of DGT~(Sec.~\ref{sec:apps}).

\section{Two-Player Two-Strategy Symmetric Games}
\label{sec_two_player}

In the simplest version of two-player two-strategy symmetric games~\cite{Hofbauer1998} individuals have a choice of two actions: \textit{cooperate} $(C)$ or \textit{defect} ($D$). Depending on their combined actions they each receive a payoff.  Since the payoffs are symmetric, we can write the full set of possible payoffs as a single payoff matrix using the convention that entries indicate the payoff received by the player whose actions occupy the rows:

\begin{equation}
  G:=\;\begin{array}{c|cc}
 & C & D \\ 
\hline
C & R  & S \\
D & T & P \\ \end{array} 
\label{gameO}
\end{equation} 
In this standard notation: both players receive the \textit{reward} $R$ if they both cooperate; both get the \textit{punishment} $P$ if they both defect; and a defector receives the \textit{temptation} $T$ when playing against a cooperator, who gets the \textit{sucker's payoff} $S$. The relative payoff values $R, S, T, P\in\mathbb{R}$ determine the nature of the game. We can standardize the payoffs by setting $R=1$ and $P=0$ and parameterize games by the $T$ and $S$ payoffs. 

\begin{figure}
\includegraphics[width=\columnwidth]{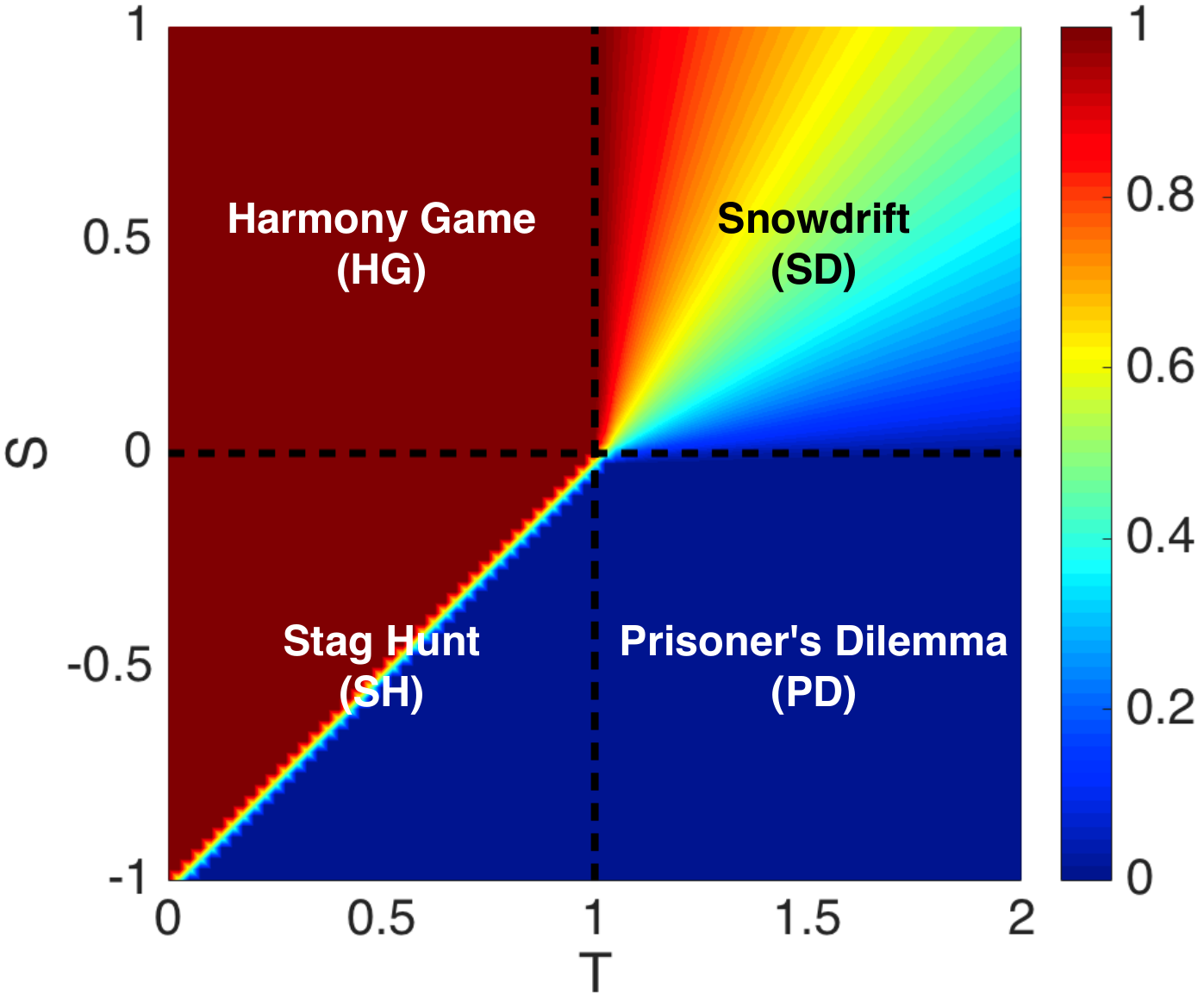}
 \caption{\textbf{T-S space}. The four classes of two-player two-strategy symmetric games within the standard $T$--$S$ parameter space ($R~=~1, P~=~0$). Prisoner's Dilemma (PD), Snowdrift (SD), Stag-Hunt (SH), and Harmony Game (HG). Colors show the level of cooperation obtained by the replicator dynamics at the stable equilibrium starting from an initial fraction of cooperators $x_0=0.5$.}
 \label{fig_TS}
\end{figure}

Figure~\ref{fig_TS} shows how we can classify games into four characteristic types according to where they lie on in the standard $T$-$S$ parameter space~\cite{perc2010coevolutionary}.  
Within such games, there may exist one or more Nash equilibria (NE)~\cite{nash1950equilibrium} --- steady states in which no player can benefit by changing strategies while the other player keeps theirs unchanged~\cite{Hofbauer1998}. 
The \textit{Prisoner's Dilemma}~(PD) game corresponds to the lower right quadrant where $T\!>\!R\!>\!P\!>\!S$. In this game defection is the rational choice such that mutual defection $(D, D)$ is the unique NE. 
In the upper right quadrant we have the \textit{Snow Drift} (SD) game in which  $T\!>\!R\!>\!S\!>\!P$. Players have an incentive to play $D$ but mutual defection is harmful for both parties. 
In the \textit{Stag Hunt} (SH) game (lower left quadrant), the payoff ordering is $R\!>\!T\!>\!P\!>\!S$, which makes mutual cooperation $(C, C)$ a NE in which both players earn the most. The SH game also contains a second NE when both players defect $(D, D)$, but results in a less favorable outcome.
Finally, in the upper left quadrant is the \textit{Harmony} game (HG), defined for $T<R$ and $S>P$, which has a single NE and payoff-dominant outcome of $(C,C)$. For more details we refer the reader to~\cite{Hofbauer1998}. 
 
Nash equilibria represent the expected behavior of rational players.  However, rather than focus on rational individuals, we instead consider a population of individuals that learn socially through processes of imitation~\cite{laland2004social}. Individuals interact with each other and can stochastically imitate their partner's strategy with a probability proportional to the difference of their payoffs. 
Specifically, at time $t$, player $i$ with strategy $s_i^{(t)}$ will imitate player $j$'s strategy with a probability that is a function of the difference in payoffs
$\pi_j^{(t)} - \pi_i^{(t)}$, where $\pi_i^{(t)}$ represents $i$'s payoff at time $t$.
If we assume an infinite and well-mixed population, the evolution of strategies can be modeled at the population level according to the proportion of cooperators $x$. 
This yields the replicator equation: 
\begin{equation}
\dot{x}=x(1-x)(\Pi_C-\Pi_D) \enspace,
\label{replic}
\end{equation}
where $\Pi_C$ and $\Pi_D$ represent the expected payoff of a cooperator and a defector, respectively, when a fraction $x$ of the population are cooperators. 
The fixed points of the replicator equation, i.e., the solutions of $\dot{x}=0$, represent the equilibria of the game dynamics.

\section{Individual perceptions in games}
\label{sec_drunk_games}
All individuals in standard two-player two-strategy symmetric games play the same game and receive the same set of payoffs given a particular set of actions played.
Here we introduce the notion of \textit{drunk games} where players may individually perceive different payoffs for the same set of actions.  
We model the simplest setting of two possible perceptions by coupling two different games $G_1$ and $G_2$, each representing a state of perception. 
In the following, we describe an example of such a game, which we call the \textit{Pub Dilemma}.

\subsection{The Pub Dilemma}
\label{sec_pub_dilemma}
In the Pub Dilemma, two individuals approach the bar of a busy pub. To receive their drinks efficiently, they decide to combine their orders, but do not discuss who will make the order and settle the bill. Both individuals attract the attention of different bar tenders simultaneously and therefore have two available actions: cooperate $C$, by offering to buy a round (buy two beers, one for each), or defect $D$, by doing nothing and hoping that the other will make the order. The payoffs are calculated as a function of the total beer $b_T$ and amount of free beer $b_F$ received. Note that for convenience we set $b_T$ to half the number of beers received to keep within the standard setting where $R=1$ and $P=0$:
\begin{equation}
b_T  :=\;\begin{array}{c|cc}
 & C & D \\ 
\hline
C & 1  & \frac{1}{2} \\
D & \frac{1}{2} & 0 \\ \end{array} \qquad
b_F  :=\;\begin{array}{c|cc}
 & C & D \\ 
\hline
C & 0  & -1 \\
D & 1 & 0 \\ \end{array} 
\label{beers}
\end{equation}

At each round, each player perceives the interaction from either a \textit{sober} state, with payoffs $G_1=b_T+b_F$, or an \textit{intoxicated} state, with payoffs $G_2=b_T$,
\begin{equation}
G_1  :=\;\begin{array}{c|cc}
 & C & D \\ 
\hline
C & 1  & -\frac{1}{2} \\
D & \frac{3}{2} & 0 \\ \end{array} \qquad
G_2  :=\;\begin{array}{c|cc}
 & C & D \\ 
\hline
C & 1  & \frac{1}{2} \\
D & \frac{1}{2} & 0 \\ \end{array} 
\label{PD_HG_payoffs}
\end{equation}
The sober perception of payoffs includes the cost of the beer and results in a PD scenario. The intoxicated individual, on the other hand, is no longer concerned with the cost and so perceives a payoff proportional to the number of beers received $b_T$, resulting in a HG scenario.

The change in perceptions between the two games is governed by an individual state variable $\alpha_i$, which we interpret as the probability for player $i$ to perceive the $G_2$ (intoxicated) game. 
After playing a round, each player updates its internal state $\alpha$ along with its strategy according to an imitation-based update rule. 
Within the pub dilemma, we define the $\alpha$-update function such that it constantly decreases over time, simulating the individual recovering to the sober state, but increases as a function of beer consumed during a round. In this way, $\alpha_i$ dynamically couples the two games such that it captures player $i$'s previous experience. We assume that the change in $\alpha_i$ is a function of both interacting players' actions, 
\begin{equation}
\dot{\alpha_i} = \kappa\alpha_i(1-\alpha_i)(b_T-\mu) \enspace,
\label{eq_ind_alpha}
\end{equation}
where the total beers $b_T$ is given in Eq.~\eqref{beers}. Parameters $\kappa$ and $\mu$ control how sensitive players' perceptions are to their prior experiences and the relative rate of decay back to the sober state, respectively.

In the same way as standard evolutionary games, players update their strategy after each round according to an imitation-based rule. 
 However, what is different to standard EGT is that the two players $i$ and $j$ may be in different states at time $t$. As such, $\pi_j^{(t)}$ is not necessarily the same as the payoff that player $i$ would obtain in the same situation. 
 In other words, when player $i$ updates their strategy they compare the $\pi_i$ and $\pi_j$ that are the payoffs as perceived by players~$i$ and $j$ respectively. 

\begin{figure}
\centering
\includegraphics[width=\columnwidth]{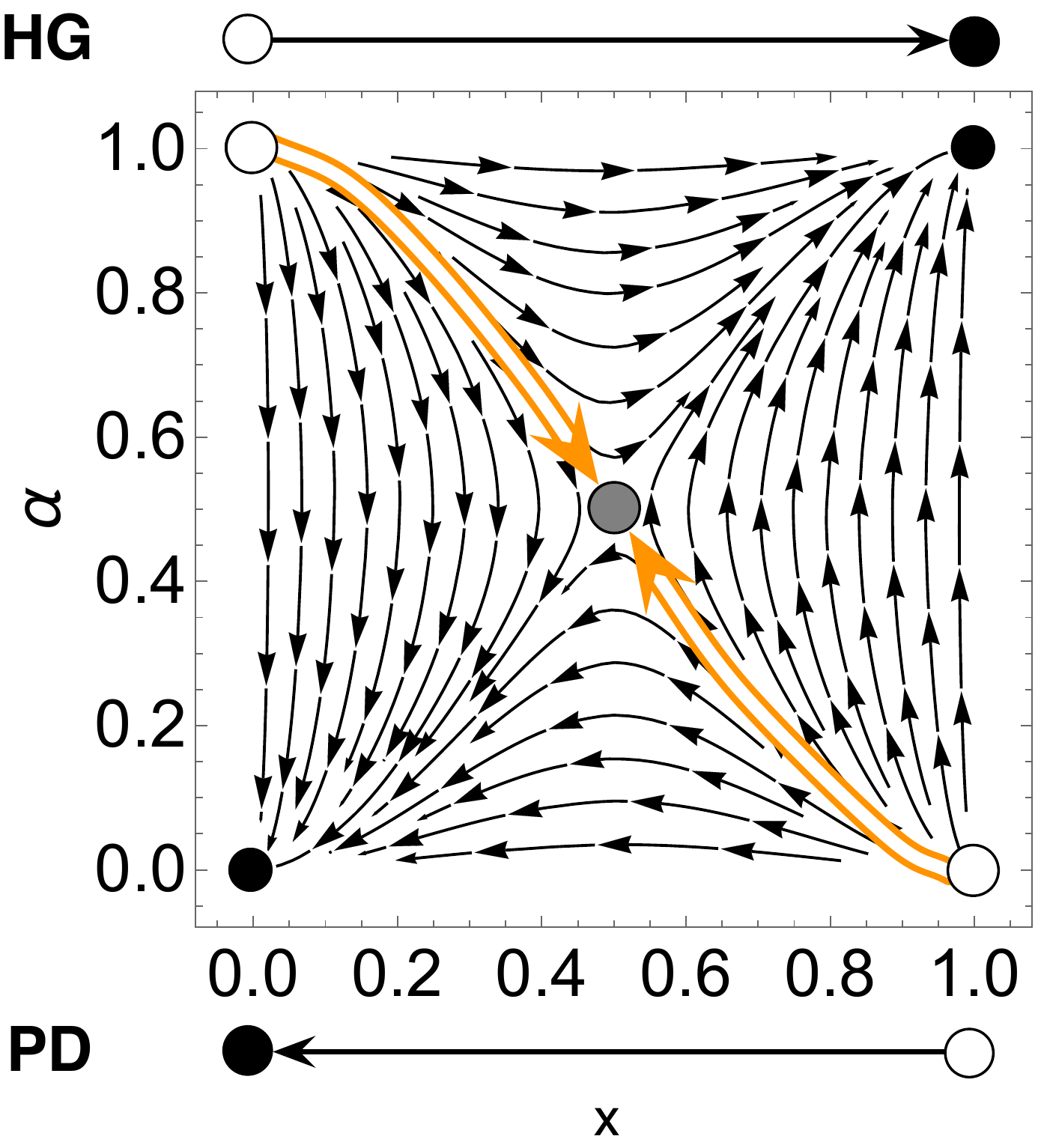}
\caption{\textbf{The Pub Dilemma}. Coupling of the Prisoner's Dilemma ($G_1$: $S_{\rm{PD}}=-0.5$, $T_{\rm{PD}}=1.5$) and the Harmony Game ($G_2$: $S_{\rm{HG}}=0.5$, $T_{\rm{HG}}=0.5$) using $\kappa=1, \mu=0.5$. The field diagram illustrates how the proportion of cooperators $x$ and the average value of $\alpha$ in the population evolves and contains a number of fixed points indicated by circle markers, which are either stable (black), unstable (white), or saddle points (gray). When $\alpha\in\{0,1\}$ we recover the game dynamics and fixed points of the original games (HG, \textit{top}; PD, \textit{bottom}). 
}
\label{fig:pubdilemma}
\end{figure}

Figure~\ref{fig:pubdilemma} illustrates the game dynamics for a population playing the Pub Dilemma for $\kappa=1$ and $\mu=0.5$ with respect to the proportion of cooperators $x$ (Eq.~\ref{replic}) and the average value of $\alpha$ (Eq.~\ref{eq_ind_alpha}). One can see that when $\alpha=0$ we recover the game dynamics of the PD game, which has a stable equilibrium at full defection. For $\alpha=1$ we obtain the HG dynamics, which has a stable equilibrium at full cooperation. The parameters of the two coupled games are symmetric with respect to the center of Fig.~\ref{fig_TS}. Consequently, the two basins of attractions have the same size, i.e., half area of the unitary $(x, \alpha)$ space. 
In addition to the fixed points of the PD and HG games, the Pub Dilemma introduces a saddle point at $x=\alpha=0.5$. This interior point is unstable and it can only be reached following the trajectories  of the orange arrows that delimit the basins of attraction for full defection and full cooperation.


\begin{figure*}
\centering
\includegraphics[width=\linewidth]{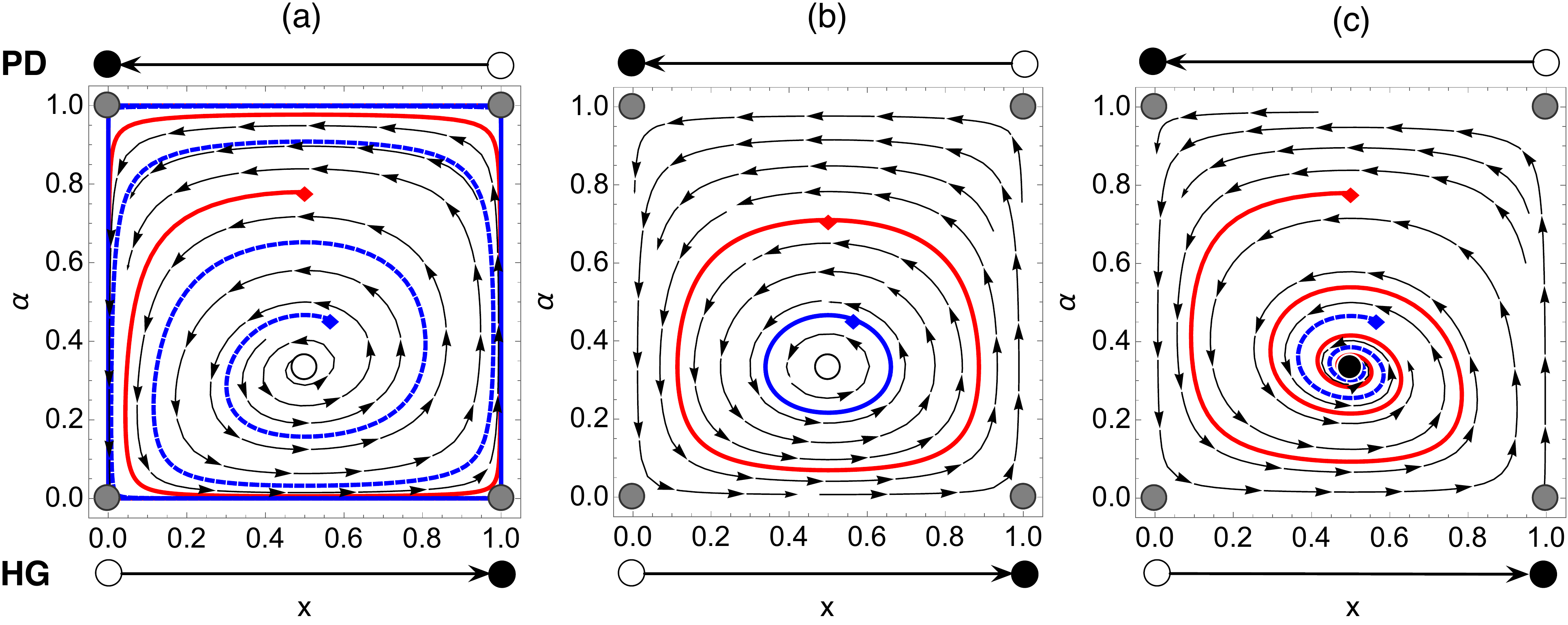}
\caption{\textbf{The Drunk Prisoner}. Coupling of a Harmony Game ($G_1$) and a Prisoner's Dilemma ($G_2$) displays a Hopf bifurcation regulated by the payoffs $S_{\rm{HG}}$ and $T_{\rm{HG}}$ ($S_{\rm{PD}}=-1$, $T_{\rm{PD}}=2$, $\kappa=1$, $\mu=0.5$ for each panel). Two trajectories (red and blue curves) for arbitrary initial conditions are shown in each example. (a): $S_{\rm{HG}}=T_{\rm{HG}}<0.5$, the Drunk Prisoner contains no stable fixed points and displays an unstable spiral originating at $x=\mu=0.5$. All spirals converge to a trajectory that follows the boundaries of the $(x,\alpha)$ plane such that $0<\alpha<1$. 
(b): the onset of the Hopf Bifurcation occurs at $S_{\rm{HG}}=T_{\rm{HG}}=0.5$ and we observe the same unstable fixed point at $x=\mu=0.5$ surrounded by an infinite set of closed cyclic trajectories. (c): $S_{\rm{HG}}=T_{\rm{HG}}>0.5$, the cycles collide over the fixed point at $x=\mu=0.5$ which becomes stable. In this setting a spirals appear such that any initialization of the system, except those on the boundaries, converge to the same interior fixed point where half of the population are cooperators and the other half defectors.
}
\label{fig:hopf}
\end{figure*}

\subsection{Drunk games}
\label{sec_gen_drunk_games}

The Pub Dilemma describes a particular coupling of games, however the same idea can be applied more generally to couple any pair of games using the state variable~$\alpha$. We describe this general formalism as a \textit{drunk game} using the notation $G_1\oplus_\alpha G_2$. At the population level, we can represent the system dynamics as:
\begin{align}
\dot{\alpha} = & \;f(x,\alpha) \label{alphagames2} \\
\dot{x} = & \;x(1-x)(\Pi_{C}-\Pi_{D})  \enspace,
\label{alphagames1}
\end{align}
in which we denote the population mean state as $\alpha$ and its evolution as a function of its current value [Eq.~\eqref{alphagames2}] and the current proportion of cooperators in the population~$x$ [Eq.~\eqref{alphagames1}]. 
The latter evolves according to the relative difference in expected payoff for the two strategies $\Pi_C$ and $\Pi_D$. However, the expected payoffs are dependent on $\alpha$ and they can be calculated as a convex combination of the two perceptions:
\begin{align}
\Pi_C & =\alpha(xR_1  + (1-x)S_2)+(1-\alpha)(xR_2 + (1-x)S_1) \notag \\
\Pi_D & =\alpha(xT_2+ (1-x)P_1)+(1-\alpha)( x T_1+ (1-x)P_2 )   
\label{eq_pi_alpha}
\end{align}
where $\{R_g,S_g,T_g,P_g\}$ are payoffs related to the game $G_g$, $g \in \{1,2\}$. Unless otherwise stated, we set $R_g=1$ and $P_g=0$. This framework does not place any requirement on the functional form that $f(x,\alpha)$ takes, as long as it satisfies the constraint $\alpha \in [0,1]$. 
Setting $\alpha~=~0$ or $\alpha~=~1$ reduces the game to the standard games $G_1$ or $G_2$, respectively. 
When $f(x,\alpha)=0$ we recover the mixed games considered in previous studies in which a fixed proportion $\alpha$ of the population plays one game while the rest of the population plays another~\cite{amaral2016evolutionary}.

\section{Dynamics of drunk games}
\label{sec_dynamics_drunk_games}
We analyze the dynamics of drunk games in terms of the fixed points that represent the equilibria of the system and their basins of attraction.
The set of fixed points of a drunk game $G_1\oplus_\alpha G_2$ includes the fixed points of both $G_1$ and $G_2$ (at $\alpha=0$ and $\alpha=1$ respectively). However the stability of these fixed points may change. In addition, new fixed points may also emerge depending on the pair of games and the choice of the $\alpha$-update function [Eq.~\eqref{alphagames2}]. 
In the following we show numerical and analytical evidence for phenomena regarding the fixed points in drunk games. These include: a loss of stability in the stable fixed points in the original games, formation of new fixed points or spirals, and changes in the basins of attraction of fixed points.

Herein, we consider a variety of drunk games in which pairs of payoff matrices are coupled by an $\alpha$-update function that can be factorized as
\begin{equation}
	\dot{\alpha} =  f(x,\alpha) = \kappa\, \alpha(1-\alpha)\, q(x) \enspace ,
    \label{eq_alpha_coupling}
\end{equation}
where $\kappa$ is a positive constant and $q(x)$ a general function that only depends on $x$.
This function satisfies the boundary conditions of $\alpha\in[0,1]$.

\subsection{Stability of original fixed points}
The fixed points $\{(\tilde{x},\tilde{\alpha})\}$ of drunk games that were present in the original games $G_1$ and $G_2$ are only stable if either:
\begin{itemize}
	\item $\tilde{\alpha}=0$, $\tilde{x}$ is stable in $G_1$, and $q(\tilde{x})<0$ \, , or
	\item $\tilde{\alpha}=1$, $\tilde{x}$ is stable in $G_2$, and $q(\tilde{x})>0$ \, .
\end{itemize}

When $q(x)=(x-\mu)$, noting that the expected value of $b_T$ is equal to the proportion of cooperators $x$, we recover the system-level $\alpha$-update function equivalent to Eq.~\eqref{eq_ind_alpha}. The Pub Dilemma~$(\rm{PD}~\oplus_\alpha~\rm{HG})$ includes both the stable fixed point of the Prisoner's Dilemma $(0,0)$ and the stable fixed point of the Harmony game $(1,1)$ when $0 < \mu < 1$, as we see Fig.~\ref{fig:pubdilemma}.

Reversing the order of the games in the Pub Dilemma forms another drunk game that we call the \textit{Drunk Prisoner} $(\rm{HG}\oplus_\alpha \rm{PD})$. 
Figure~\ref{fig:hopf} illustrates  the dynamics of the Drunk Prisoner and shows that neither of the fixed points from HG or PD are stable any more.

\subsection{New fixed points and spirals}

The coupling of standard two-player games can produce additional fixed points inside the boundary of the $(x,\alpha)$-plane, i.e., interior fixed points $\{(\tilde{x}^\bullet,\tilde{\alpha}^\bullet)\}$ such that $0 < \tilde{x}^\bullet < 1$ and $0 < \tilde{\alpha}^\bullet < 1$. 
To analyze these interior fixed points, we first rewrite the cooperation dynamics in Eq.~\eqref{alphagames1} by substituting $\Pi_C, \Pi_D$ for the expressions in Eq.~\eqref{eq_pi_alpha}:
\begin{equation}
\dot{x}=-x(1-x)\left[(1-\alpha)\,h_1(x)+\alpha\,h_2(x)\right],
\label{xdoteq}
\end{equation}
where $h_g(x)=(1-x)\mathcal{F}_g+x\mathcal{G}_g$ represents the incentive to defect in game $g$ given the current proportion of cooperators $x$, i.e., the \textit{fear} of cooperating $\mathcal{F}_g = P_g-S_g$ when your opponent defects and the \textit{greed} $\mathcal{G}_g = T_g-R_g$ from the possibility of exploiting your opponent's cooperation~\cite{macy2002learning}. 
In order for one of these interior points $(\tilde{x}^\bullet,\tilde{\alpha}^\bullet)$ to be a fixed point, $q(\tilde{x}^\bullet)$ and $\dot{x}$ must be equal to zero. Then 
from Eq.~\eqref{xdoteq} we obtain: 
\begin{eqnarray}
\tilde{\alpha}^\bullet = \frac{h_1(\tilde{x}^\bullet)}{h_1(\tilde{x}^\bullet) - h_2(\tilde{x}^\bullet)} \enspace,
\label{alpha0}
\end{eqnarray}
which implies that  $h_1(\tilde{x}^\bullet)$ and $h_2(\tilde{x}^\bullet)$ must have different signs to ensure that $0<\tilde{\alpha}^\bullet<1$.

We can determine the stability of this interior fixed point using the eigenvalues of the Jacobian of the system in Eq.~\eqref{alphagames2} and Eq.~\eqref{alphagames1} evaluated at $(\tilde{x}^\bullet,\tilde{\alpha}^\bullet)$. The eigenvalues can be written in the form of $\lambda= u\pm i\sqrt{v}$ such that
\begin{equation}
u=\frac{\tilde{x}^\bullet(1-\tilde{x}^\bullet)\tilde{\alpha}^\bullet}{2} \,
\frac{\mathcal{F}_{2}\mathcal{G}_{1}-\mathcal{F}_{1}\mathcal{G}_{2}}{h_1(\tilde{x}^\bullet)} \enspace ,
\label{Ralpha0}
\end{equation}
and
\begin{equation}
v=\kappa \, \tilde{x}^\bullet (1-\tilde{x}^\bullet)\,\tilde{\alpha}^\bullet h_2(\tilde{x}^\bullet)q'(\tilde{x}^\bullet)-u^2  \enspace ,
\label{Ialpha0}
\end{equation}
where $q'(\tilde{x}^\bullet)$ is the derivative of $q(x)$ with respect to $x$ evaluated at $\tilde{x}^\bullet$.
When $v<0$ the eigenvalues are real (i.e., when $h_2(\tilde{x}^\bullet)q'(\tilde{x}^\bullet)<0$). We know that a pair of negative real eigenvalues indicate that a fixed point is stable~\cite{grobman1959homeomorphism,hartman1960lemma}. Therefore, when the eigenvalues are real, the interior fixed point is stable if and only if $u<0$ and the first term in the right hand of Eq.~\eqref{Ialpha0} is positive, i.e., $h_2(\tilde{x}^\bullet)q'(\tilde{x}^\bullet)>0$. 
However, if $v>0$ the eigenvalues are complex conjugates of each other.
In this case the dynamics form spirals around the interior resting point, which is either an attractor when $u<0$ or a repeller when $u>0$. In the special case when $u=0$, these orbits become limit cycles.
In summary:
\begin{enumerate}
	\item {An interior fixed point $(\tilde{x}^\bullet, \tilde{\alpha}^\bullet)$}     exists if $0<\tilde{x}^\bullet<1$, $q(\tilde{x}^\bullet)=0$, and $h_1(\tilde{x}^\bullet)h_2(\tilde{x}^\bullet)<0$.
	\item Spirals are formed if $v > 0$. The spirals are attractive when $u<0$ and repellent when $u>0$. Limit cycles are formed for the special case $u=0$.
	\item No spirals are formed if $v < 0$. Then the fixed point is stable if $\lambda < 0$ and unstable otherwise.
\end{enumerate}

We can observe these dynamics in play in the Pub Dilemma in Fig.~\ref{fig:pubdilemma} and in the Drunk Prisoner in Fig.~\ref{fig:hopf}. In the Pub Dilemma $(\rm{PD}\oplus_\alpha \rm{HG})$ with $q(x)=(x-\mu)$ a fixed point occurs at $x=\mu$. Since $h_{\rm{PD}}(x)>0$ and $h_{\rm{HG}}(x)<0$ for any value of $x$, $\lambda$ is strictly non-negative and so the interior fixed point of the Pub Dilemma is always unstable. 
However, in the Drunk Prisoner $(\rm{HG}\oplus_\alpha \rm{PD})$ 
we observe a more diverse range of game dynamics.  Now $v$ may be positive or negative and so we can observe the full range of cases given in conditions 2 (spirals are formed) and 3 (no spirals) above.  

Figure~\ref{fig:hopf} shows a set of examples of the Drunk Prisoner in which we vary the payoffs $(S_{\rm{HG}},\;T_{\rm{HG}})$ of the sober state, while keeping the payoffs of the intoxicated state fixed ($S_{\rm{PD}}=-1,\;T_{\rm{PD}}=2$). For each of these games $v>0$ and so the game dynamics exhibit\tb{s} spirals around the interior fixed point. When $S_{\rm{HG}}=T_{\rm{HG}}<\mu$ (Fig.~\ref{fig:hopf}a), $u$ is positive and the interior fixed point is unstable. When $S_{\rm{HG}}=T_{\rm{HG}}>\mu$ (Fig.~\ref{fig:hopf}c), $u$ is negative and the interior fixed point attracts all trajectories initialized anywhere other than the four extremal saddle points. In the case that $S_{\rm{HG}}=T_{\rm{HG}}=\mu$ (Fig.~\ref{fig:hopf}b), a Hopf bifurcation occurs creating an unstable fixed point surrounded by closed cycles. In terms of the pub metaphor, the population playing this particular Drunk Prisoner game will, on average, experience an endless cycle of cooperating, getting drunk, defecting, and sobering up.  

More generally, when {$S_{\rm{HG}}\neq T_{\rm{HG}}$}, the Drunk Prisoner's interior fixed point is stable and attractive when the following condition is satisfied:
\begin{equation}
  \frac{\mathcal{F}_{\rm{HG}}}{\mathcal{G}_{\rm{HG}}}>\frac{\mathcal {F}_{\rm{PD}}}{\mathcal{G}_{\rm{PD}}} \enspace .
\end{equation}
In other words, the interior fixed point becomes attractive when the fear-greed ratio is higher in the HG than in the PD game. By setting $R_g=1$ and $P_g=0$ we obtain:
\begin{equation}
 \frac{S_{\rm{HG}}}{1-T_{\rm{HG}}} > \frac{S_{\rm{PD}}}{1-T_{\rm{PD}}}\enspace .
\end{equation}

 \begin{figure*}[t]
\centering
\includegraphics[width=\linewidth]{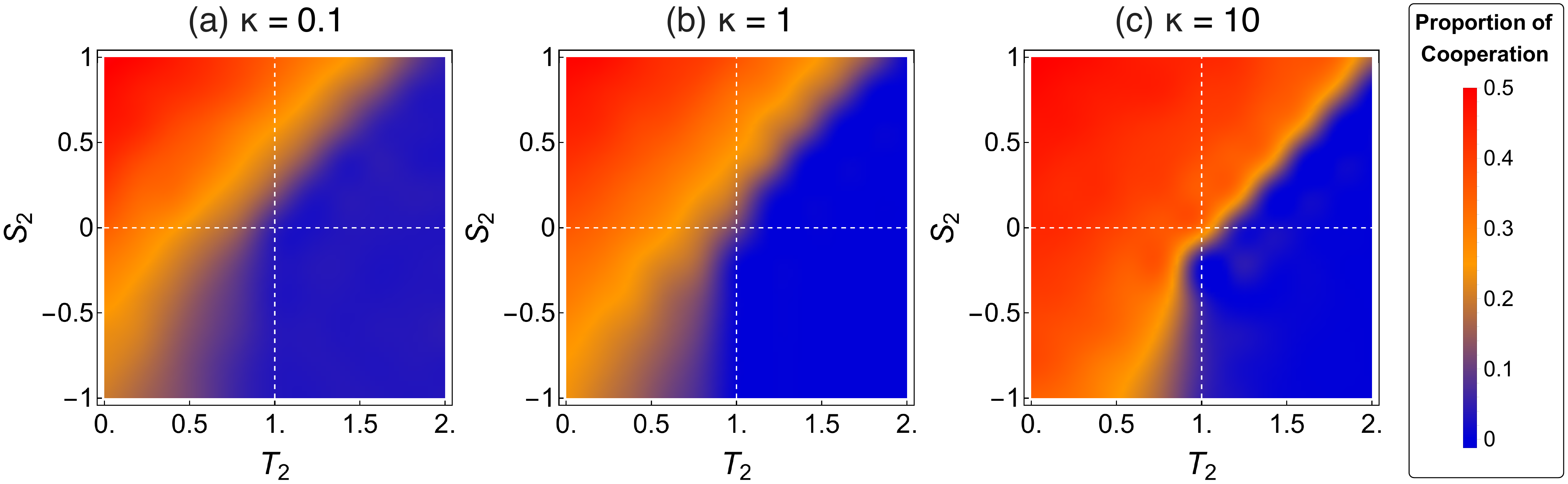}
\caption{\textbf{Generalized Pub Dilemma}. Attractiveness of the cooperation basin for the generalized Pub Dilemma: coupling the PD ($G_1$: $S_1=-1,T_1=2$) with another game $G_2$ having parameters $T_2$ and $S_2$.  The $(S_2, T_2)$ space indicates the probability of converging on the cooperative fixed point ($x=1$ and $\alpha=1$), i.e. the proportion of cooperation of the coupled games, when $\kappa=0.1$ (a), $\kappa=1$ (b) and $\kappa=10$ (c). As $\kappa$ increases we can see an overall increase of the attractiveness of cooperation. 
}
\label{k-magnitude}
\end{figure*}

\begin{figure*}
\centering
\includegraphics[width=\linewidth]{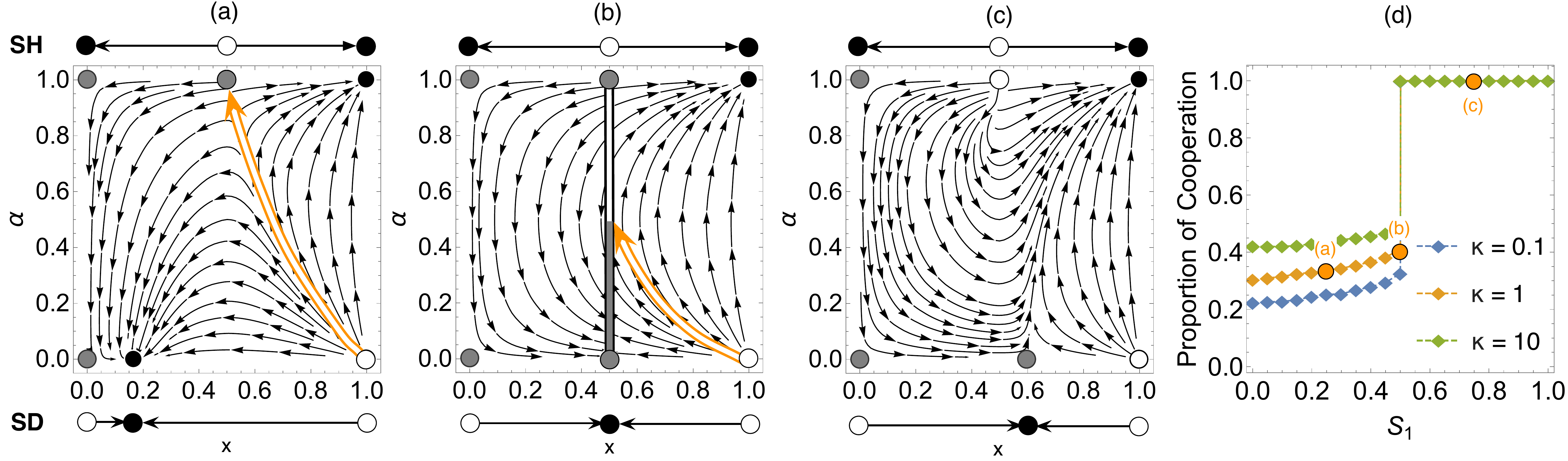}
\caption{\textbf{The Drunken Battle of Coordination}. Coupling of a Snowdrift ($G_1$: $T_{\rm{SD}}=2$, $S_1=S_{\rm{SD}}$) and the Stag Hunt ($G_2$: $T_{\rm{SH}}=0.5$, $S_{\rm{SH}}=-0.5$). (a) $S_{\rm{SD}}=0.25$ and $\kappa=1$, all trajectories converge on either the stable fixed point of SD or the stable fixed point of SH at full cooperation, with the exception of points initialized at a fixed point or on the orange arrow. (b) $S_{\rm{SD}}=0.5$ and $\kappa=1$, a first-order phase transition occurs and a line of fixed points emerge at $x=0.5$. (c) $S_{\rm{SD}}=0.75$ and $\kappa=1$, all trajectories converge on the full cooperation fixed point at the top right. (d) Attractiveness of the cooperation basin for all possible $S_1\in[0,1]$; (a)--(c) cases are marked with $\kappa=1$. Increasing $\kappa$ increases the overall attractiveness of cooperation. A first-order phase transition occurs at $S_1=0.5$ such that when $S_1>0.5$ all initial conditions lead to full cooperation.}
\label{fig:battle}
\end{figure*}

\subsection{Attractiveness of cooperation}
In standard two-player two-strategy symmetric games we can examine the attractiveness of fixed points by studying the relative size of the set of initial conditions (proportion of cooperators $x$) that eventually converge upon a particular fixed point. The attractiveness provides an estimate of the size of the basin of attraction. In drunk games, these basins of attraction are defined over the $(x,\alpha)$ plane rather than just on the $x\in[0,1]$ line. The size and shape of the basins depend on the $\alpha$-{update} function as well as on $G_1$ and $G_2$ parameters. 

We now examine how changes in the basins of attraction occur in  a generalized version of the Pub Dilemma in which $G_1$ is a Prisoner's Dilemma (with $S_1=-1$ and $T_1=2$) and $G_2$ is another game ($S_2 \in [-1,1]$ and $T_2 \in [0,2]$). 
Assuming $q(x)=(x-0.5)$, we estimate the basins of attraction for any given set of game parameters using Monte Carlo simulations.  We calculate the attractiveness of cooperation by counting the proportion of $10^3$ independent simulations
that converge upon full cooperation with initial conditions $(x_0,\alpha_0) \in [0,1]^2$ sampled uniformly at random.   

Figure~\ref{k-magnitude} shows the proportion of simulations that converge to a full cooperation fixed point for different settings of the $G_2$ payoffs $\{S_2,T_2\}$ and for different values of $\kappa$. We see that by increasing $\kappa$ the overall attractiveness of cooperation increases.  The maximal attractiveness of cooperation occurs when half of all initial conditions converge on full cooperation. Also, note that coupled games in the bottom right quadrant cannot converge on full cooperation as they correspond to the set of drunk games in which both games are versions of the PD, for which the evolutionary stable strategy is full defection.

\subsection{Phase transition in cooperation attractiveness}
We formulate and extend our previous analysis to the \textit{Drunken Battle of Coordination}, a combination of Snow-Drift and Stag Hunt games {($SD \oplus_\alpha SH$)}.  Recall that Nash equilibria in both games require the coordination of both players, i.e., in SD games the NE occurs when players choose different strategies, while in SH games a NE requires players to play the same strategy.

Figure~\ref{fig:battle} shows all dynamics in the Drunken Battle of Coordination in which we fix three of the payoff parameters ($T_{\rm{SD}}=2,S_{\rm{SH}}=-0.5,T_{\rm{SH}}=0.5$) as we vary the sucker's payoff in the SD game ($S_1=S_{\rm{SD}}~\in~[0,1]$). We continue using the same $\alpha$-update function as before with $q(x)=(x-0.5)$. Similar to the generalized Pub Dilemma, we see in Fig.~\ref{fig:battle}d that increasing $\kappa$ has the effect of increasing the attractiveness of cooperation. However, in contrast to the generalized Pub Dilemma, the Drunken Battle of Coordination displays a discontinuous transition in the attractiveness of cooperation as we vary $S_1$. Specifically, we see that a first-order transition occurs at the critical value of $S_1=0.5$ and for any $S_1$ above this value all initializations lead to full cooperation.  
The other panels in Figure~\ref{fig:battle} provide more detail, showing the dynamics in the $(x,\alpha)$ plane for three settings: (a) $S_{\rm{SD}}=0.25$, (b) $S_{\rm{SD}}=0.5$, and (c) $S_{\rm{SD}}=0.75$; also labeled in panel (d).  We see in all three cases that the stable fixed point of the SH that corresponds to complete defection becomes unstable, while the full cooperation fixed point remains stable. The stable fixed point of the SD only remains stable when $S_{\rm{SD}} \leq 0.5$ (Fig.~\ref{fig:battle}a). At the critical point, when $S_{\rm{SD}} = 0.5$ (Fig.~\ref{fig:battle}b), a line of unstable interior fixed points appears.  These fixed points are stable with respect to $x$ for values of $\alpha<0.5$.  When  $S_{\rm{SD}} > 0.5$ (Fig.~\ref{fig:battle}c), the stable fixed point of the SH game moves to the right of $x=0.5$ and becomes unstable.  As a consequence, all initializations, except those on a fixed point, 
converge to the full cooperation fixed point in the top right corner.

\begin{figure*}
\centering
\includegraphics[width=\linewidth]{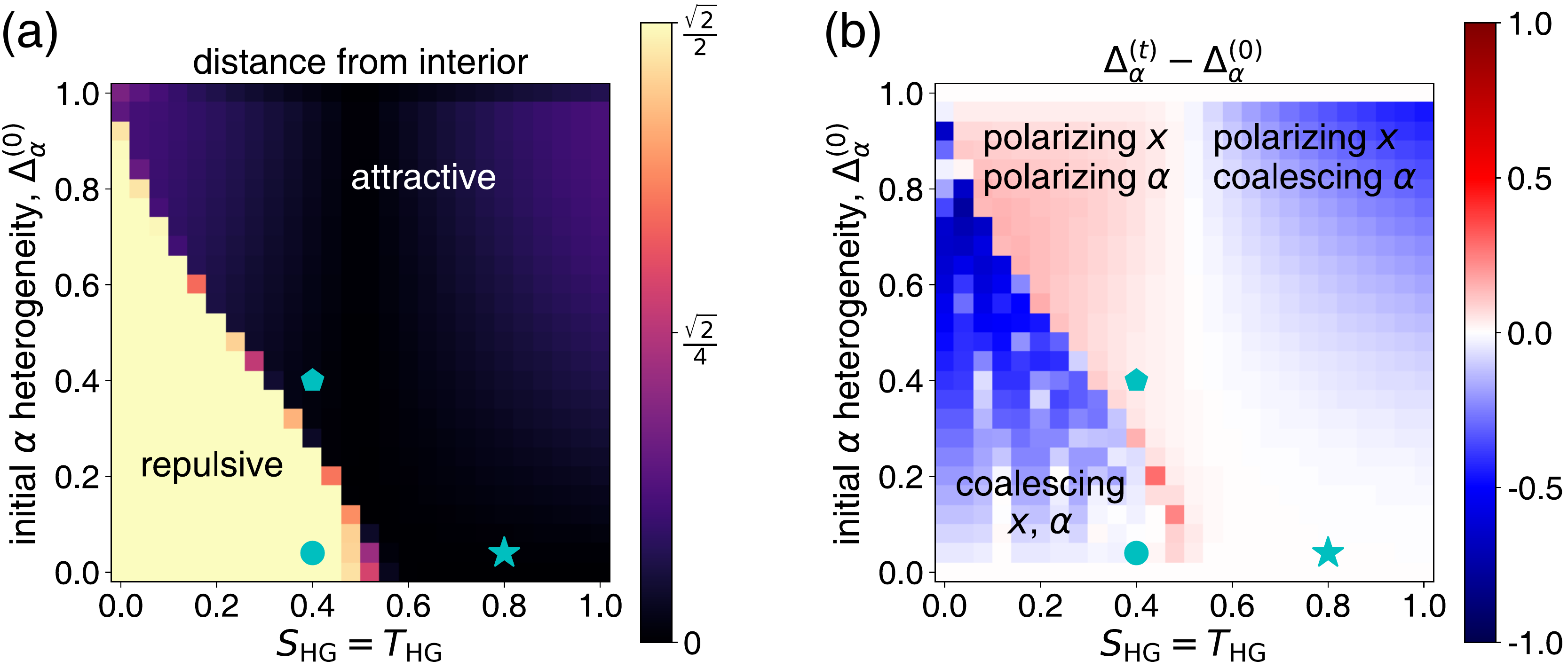}
\caption{\textbf{$\alpha$-heterogeneity in the Drunk Prisoner game}. The effect of initializing the Drunk Prisoner game with different payoffs, $S_{\rm{HG}}=T_{\rm{HG}}$, (\textit{x-axis}) and levels of $\alpha$-heterogeneity, $\Delta_{\alpha}^{(0)}$, (\textit{y-axis}). (a) heatmap of the distance of the population average from the interior fixed point. Increasing the initial $\alpha$-heterogeneity increases the stability of the interior fixed point and the Hopf bifurcation occurs at lower payoffs $S_{\rm{HG}}$ and $T_{\rm{HG}}$. (b) heatmap of the change in $\alpha$-heterogeneity, $\Delta_{\alpha}^{(t)}-\Delta_{\alpha}^{(0)}$, at time $t=10^4$. Three distinct behaviors are observed according to whether the strategies, $x$, and perceptions, $\alpha$ coalesce or polarize. The cyan markers indicate the individual trajectories shown in Figure~\ref{fig:ABM}: circle (a), star (b) and pentagon (c). }\label{fig:comparison}
\end{figure*}

\section{The effect of heterogeneous perceptions}
\label{sec_agent_based}

So far we have made a mean-field approximation by assuming that the system of individuals can be represented by the population averages. We now consider the dynamics of the system when we introduce populations of individuals with different perceptions, i.e., there is some variance in the individual $\alpha_i$ values.  Modeling the evolution of individual perceptions becomes analytically intractable and so we use an agent-based model (ABM) to simulate the interactions of a large population.  
We restrict our current investigation to the simple setting in which individuals start with one of two possible perceptions, $\alpha_{i}^{(0)} = \{\alpha_{\mathbf{1}}, \alpha_{\mathbf{2}}\}$ and $\alpha_{\mathbf{1}} < \alpha_{\mathbf{2}}$, such that individuals initialized with $\alpha_{\mathbf{1}}$ have a greater initial propensity to perceive $G_1$.  We define the heterogeneity of $\alpha$ as:
\begin{equation}
	\Delta_{\alpha} = \frac{\alpha_{\mathbf{1}} - \alpha_{\mathbf{2}}}{\alpha_{\mathbf{1}} + \alpha_{\mathbf{2}}} \enspace ,
\end{equation}
such that $\Delta_{\alpha} = 0$ indicates that $\alpha$ is homogeneous and $\Delta_{\alpha} = 1$ means that $\alpha_{\mathbf{1}} = 0$ and $\alpha_{\mathbf{2}} = 1$. 
In this binary setting the mean behavior is no longer representative of any of the individuals in the population. 

At each round of the ABM simulation each of the $N$ agents play the drunk game with every other agent ($N(N-1)/2$ games are played each round), accumulate payoffs according to their actions and the game they perceive ($G_1$ with probability $1-\alpha_i^{(t)}$ and $G_2$ otherwise). After each round the strategies $\{x_i\}$ and perceptions $\{\alpha_i\}$ are updated synchronously such that every agent's strategy and perception is updated at time $t+1$ according to the agent strategies and perceptions at time $t$. 

To minimize confounding effects, we set up the ABM to match the analytical setting as closely as possible. For instance, to minimize finite-size effects, we use a relatively large population of $N=10^4$ agents. Agents have pure strategies, either cooperate $(C)$ or defect $(D)$ that are initialized randomly according to $\textrm{Pr}[x_i^{(0)} = C] = x^{(0)}$.  All agents update their $\alpha_i$ according to (cf.~Eq.~\eqref{eq_ind_alpha}):
\begin{equation}
\alpha^{(t+1)}_{i}=\alpha^{(t)}_{i} + \kappa \alpha^{(t)}_{i}(1-\alpha^{(t)}_{i}) (\bar{x}^{(t)} - \mu) \enspace,
\end{equation}
where $\bar{x}^{(t)}$ is the proportion of cooperators in the population at time $t$.
Agents update their strategy according to the local replicator rule~\cite{roca2009}.  In the local replicator rule, each agent $i$ randomly chooses another agent $j$ and imitates $j$'s strategy for the next round $(t+1)$ with probability $p_{ij}^{(t+1)}$ given by:
\begin{equation}
	p_{ij}^{(t+1)} = \max\left(0, \beta\frac{\pi_j - \pi_i}{\Phi} \right) \enspace ,
\end{equation}
where $\pi_i^{(t)}$ is the average payoff agent $i$ receives in round $t$ and $\Phi$ is the maximum possible difference in payoffs, i.e., $(N-1)[\max(1,T)-\min(0,S)]$. The parameter $\beta<1$ controls the intensity of selection and thus the update strategy change rate in the system. We set $\beta=\kappa=0.1$ to enact a gradual change and to achieve greater numerical stability in finite size populations.

For all the games presented so far, when $\Delta_{\alpha} = 0$ we find that the system behavior matches the results of the analytical ones, all agents follow the same trajectory until they meet one of the stable fixed points. In many of the games, introducing heterogeneity ($\Delta_{\alpha} > 0$) often has little effect on the final outcome, but can increase the time scale for agents to converge upon a stable fixed point. 
A more substantial effect of heterogeneity occurs in games that contain a stable interior fixed point. To better show this phenomenon, we consider the Drunk Prisoner game ($\rm{HG} \oplus_\alpha \rm{PD}$) for which we previously established that an interior stable fixed point exists when $S_{\rm{HG}}=T_{\rm{HG}}>0.5$. Figure~\ref{fig:comparison}a shows the distance of $(\bar{x}^{(t)},\bar{\alpha}^{(t)})$ from the interior equilibrium averaged over the whole population at $t=10^4$. We see that when the initial $\alpha$ heterogeneity $\Delta_{\alpha}^{(0)} =0$, the Hopf bifurcation occurs at $S_{\rm{HG}} = T_{\rm{HG}} = 0.5$, which is in agreement with our analytical results. However, increasing $\Delta_{\alpha}^{(0)}$, we find that the bifurcation occurs at lower values of   $S_{\rm{HG}}$ and $T_{\rm{HG}}$. Put differently, the stability of the interior fixed point increases as the heterogeneity of perceptions increases, at least when we consider the average over the whole population.

\begin{figure}
\includegraphics[width=\linewidth]{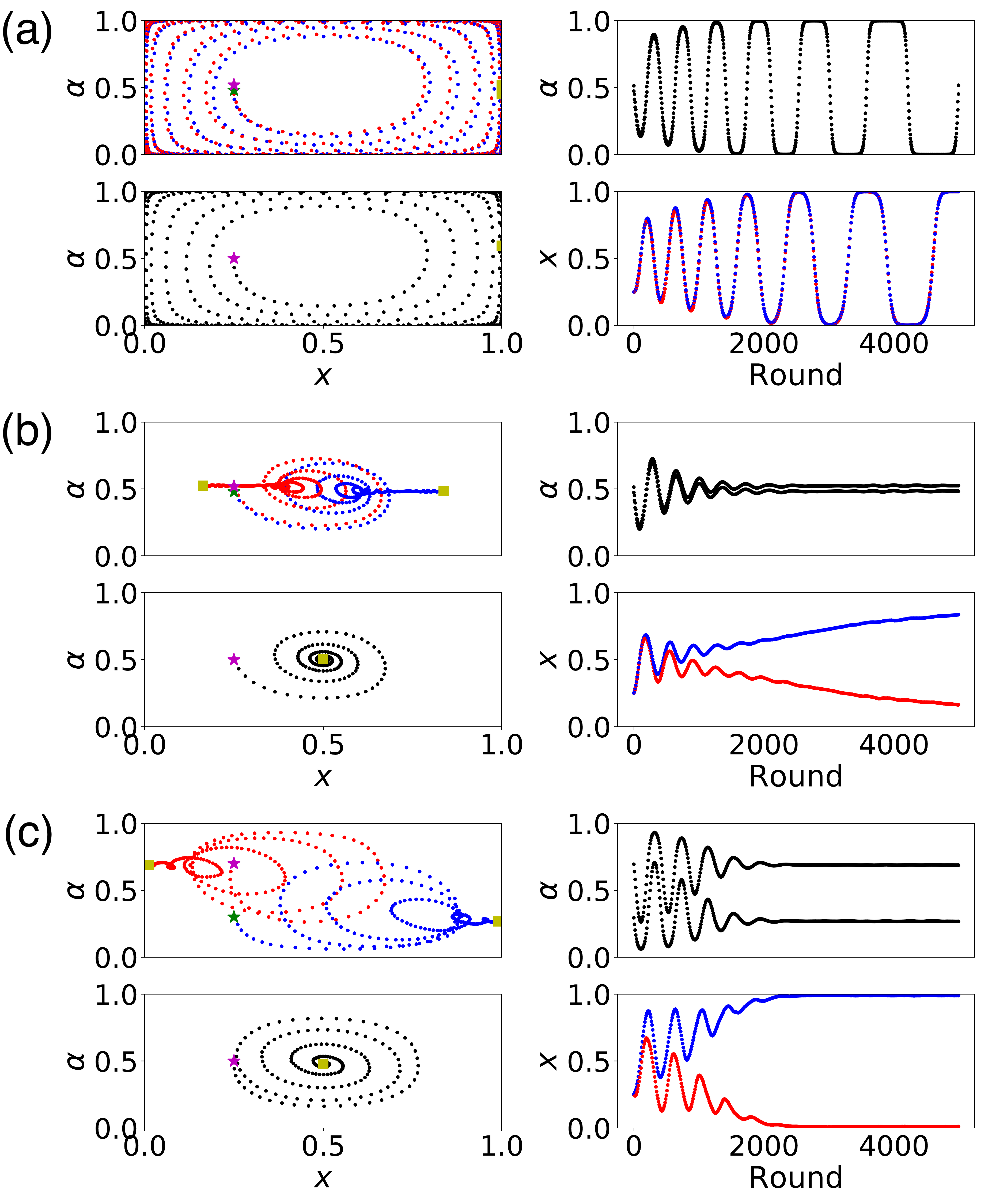}
  \caption{\textbf{ABM simulations of the Drunk Prisoner game}. Each of the panels (a)--(c) displays four plots describing the dynamics of the proportion of cooperators $x$ and average perception $\alpha$ over $t=5 \times 10^3$ rounds. The left plots show the trajectory through the $(x,\alpha)$-plane starting from the star. The top plot shows two trajectories for each of the $\alpha$ initial values, while the bottom plot show the mean over the whole population. The plots on the right show the evolution of $\alpha$ (top) and $x$ (bottom) over time. (a) $S_{\rm{HG}} = T_{\rm{HG}} = 0.4$ and $\Delta_{\alpha}^{(0)} = 0.04$ (circle marker in Fig.~\ref{fig:comparison}), (b) $S_{\rm{HG}} = T_{\rm{HG}} = 0.8$ and $\Delta_{\alpha}^{(0)} = 0.04$ (star marker in Fig.~\ref{fig:comparison}), and (c) $S_{\rm{HG}} = T_{\rm{HG}} = 0.4$ and $\Delta_{\alpha}^{(0)} = 0.4$ (pentagon marker in Fig.~\ref{fig:comparison}).  We see that small amounts of initial heterogeneity in the population can result in very different outcomes relative to the mean-field results.}
  \label{fig:ABM}
\end{figure}

The mean of the population, however, is not representative of any of the individual agents in the population (when $\Delta_{\alpha}^{(0)}>0$) due to the bimodal distribution over~$\alpha$. Figure~\ref{fig:comparison}b displays a heat map of the change in $\alpha$ heterogeneity from the start to the end of the simulation, i.e., $\Delta_{\alpha}^{(t)} - \Delta_{\alpha}^{(0)}$. Comparing against Figure~\ref{fig:comparison}a, we see that when the interior fixed point is unstable, the strategies and perceptions coalesce to become homogeneous and converge upon a trajectory that follows the boundaries of the $(x,\alpha)$ plane. Figure~\ref{fig:ABM}a shows, for the parameter setting indicated by a circle marker in Fig.~\ref{fig:comparison}, an example of this type of trajectory. In this case there is little difference between the individual and population dynamics.
When $S_{\rm{HG}} = T_{\rm{HG}} > 0.5$ we observe that the distribution of $\alpha$ coalesces to a single mode, but the strategies polarize such that the agents initialized at $\alpha_{\mathbf{1}}$ become cooperators, while the rest become defectors. Figure~\ref{fig:ABM}b shows an example for the settings indicated by a star marker in Fig.~\ref{fig:comparison}. In this example we see that only a relatively small amount of initial heterogeneity ($\Delta_{\alpha}^{(0)}=0.04$) is required to cause this polarization of strategies. 
When $S_{\rm{HG}} = T_{\rm{HG}} < 0.5$ and the interior fixed point is, for the global system behavior, stable, we find that the strategies also diverge. However, in this regime the perceptions also polarize such that $\alpha_i$ of agents diverge according to their initial values and resulting in $\alpha_{\mathbf{1}}^{(t)} \rightarrow 0$ and $\alpha_{\mathbf{2}}^{(t)} \rightarrow 1$. Finally, Fig.~\ref{fig:ABM}c shows an example of this setting (pentagon marker in Fig.~\ref{fig:comparison}). In this case the observations appear somewhat paradoxical between the two different scales: the global behavior results in attractive spirals while the behavior at the individual level is repulsive with respect to the interior fixed point.

\section{Potential applications of Drunk Game Theory}
\label{sec:apps}

We have presented our framework in the context of social consumption of alcohol as it provides an easy-to-relate-to scenario in which perceptions may change over time, vary between individuals and change as a consequence of the outcomes of previous interactions. The analogy, however, extends to much more diverse range of systems and settings in which interacting agents may have their own individual states and these states change over time as a function of their experience.

For instance, experiments have found that in public-goods games people vary in their personal preferences for fairness, with some of them being conditional cooperators~\cite{fischbacher2001people}, i.e., cooperating more as they experience more cooperation. Conditional cooperation has recently been demonstrated to emerge through different levels of individual understanding of how to maximize income~\cite{burton2016conditional}, which we might consider as different perception states in a drunk game. In pairwise coordination games it has also been shown that repeated coordination tasks can elicit a sense of commitment in agents, reminiscent of an evolving individual state. Because of such commitment, agents change their perception of the game over time and end up cooperating more than expected, even through fluctuations of interest and trust~\cite{john2018chains}.

In addition, time-evolving, individual perception levels are relevant in the dynamics of social groups or organizations, like for instance how trust can be built or broken between interacting organizations and/or individuals~\cite{putnam1994making, kreps1982rational}, or how innovation and financial investments can alter the perceived benefits of competing technologies among individual firms~\cite{arthur1989competing}. There may be also other mechanisms for changing perceptions such as diminishing returns for repeated actions; for instance, the benefit of scoring points in team sports may change depending on whether a team is currently leading or not~\cite{peel2015predicting}. 
Also, those systems that involve some level of consensus forming are related, e.g., naming conventions in social systems~\cite{centola2018experimental} or quorum sensing in biological systems and insect populations~\cite{miller2001quorum}. In these types of systems individual states are related to the population density observed by the individual, which will modify the benefit associated with different actions. 
 Previous studies indicate that the emergence of synchronization may be a consequence of an evolutionary non-cooperative game in which individuals decide their behavior according to the state resulting from their previous interactions~\cite{antonioni2017coevolution}. 
Finally, in prebiotic biology we see analogies to memory and perception in prebiotic chemistry, where replicating RNA molecules change their conformation in response to previous interactions with other RNA molecules~\cite{yeates2016dynamics}.

\section{Discussion}
\label{sec_discussion}

In many complex systems macroscopic, critical behavior can arise from the combination of simple, local interactions among individual agents. A crucial example is the emergence of cooperation in game-theoretic settings~\cite{nowak2006evolutionary}. Previous approaches assumed homogeneous interactions both across agents and over time. Our drunk games provide a new dynamical, individualistic view on past approaches, endowing each agent with a distinct, time-evolving perception of the consequences of every interaction. In this way, two agents can engage in the same choice but experience different individual payoffs. 
Using a mean-field approximation we can analyze the behavior of the population at a macroscopic level. This approach provides us with a indication of how the population, on average, evolves over time with respect to the strategies they play and the payoffs they perceive.  This co-evolution of perspectives and strategies provides an interesting departure from the standard mixed games~\cite{cressman2000evolutionary,hashimoto2006unpredictability}, in which a given proportion of the population plays one game while the rest play another.  For unstructured populations mixed games produce a trivial result in which the level of cooperation that emerges is equal to that of a standard two-player game in which the payoffs are the average, weighted by proportion of players, of the two games. In drunk games the proportion of players that play each game changes in response to the previous outcomes. Therefore, their analysis is not so trivial anymore and it is not possible to compare against a simple weighted average of games.  

The mean-field approximation provides analytical tractability at the cost of treating each individual as an average player, which potentially may not be representative of any of the individuals in the population. In principle, by modeling all individuals with an identical, but evolving perception state is similar to the accounting for a background or environment changing state --- the mean-field results of our Drunk Prisoner game closely resemble the recently presented oscillating tragedy of the commons~\cite{weitz2016}. Our agent-based model simulations thus play a crucial role in probing the relationship between micro- and macroscopic behaviors by allowing us to capture the dynamics of the individuals in the population.  
When considering heterogeneous populations, we observe that qualitatively similar behaviors at the macroscopic scale can confound very different behaviors at the microscopic scale. We found that often highly divided initial states of perception coalesce.  This coalescence might seem unsurprising given that the population is well mixed and all players interact with every other player.  However, our results are in stark contrast to other settings where small levels of heterogeneity in initial individual perception states cause the population to polarize. 
Furthermore, even when perceptions coalesce, we find that sometimes the heterogeneous initialization can cause a complete polarization of strategies. 

The framework of drunk games opens up a number of potential avenues of investigation. We considered drunk games with only two distinct perceptions  (payoff matrices). However, the framework could easily be extended to allow for a multitude of perceptions by replacing the Bernoulli states $\alpha_i$ with a categorical probability distribution indicating the probability of playing one of $k$ games. Or to a continuously varying set of payoffs~\cite{DGT}. Recent evidence indicates that biological diversity creates differences in the way that individuals transition between cognitive states~\cite{amico2018quest}. Such a finding might motivate the exploration of a type of drunk game in which the $\alpha$-function varies between individuals.


\section{Acknowledgements}
We thank Sanja Selakovic, J-P.~Gonzales and the participants of the SFI CSSS 2014 for supporting the conception of DGT. We also thank Markus Brede, Aaron Clauset, Javier Garcia-Bernardo, Jelena Grujic, Christian Hilbe, Abigail Z.~Jacobs, Cynthia Siew, John Michael and Dawid Walentek for helpful conversations and feedback. We are grateful for the series of Winter Workshop on Complex Systems for supporting the development of DGT. Authors contributed equally to this work and appear alphabetically. All authors contributed equally to the purchase of beer. 
This work was supported by SNSF grants no.~P2LAP1-161864 and P300P1-171537 [AA], EPSRC Doctoral Training Centre grant (EP/G03690X/1)[MS], the F.R.S-FNRS and the Concerted Research Action (ARC) programme (contract ARC 14/19-060) of the Federation Wallonia-Brussels [LP], ERCIM Alain Bensoussan Fellowship Programme [LAM-V].

\newpage

\bibliography{games2.bib} 

\clearpage

\begin{figure*}
\includegraphics[width=.8\linewidth]{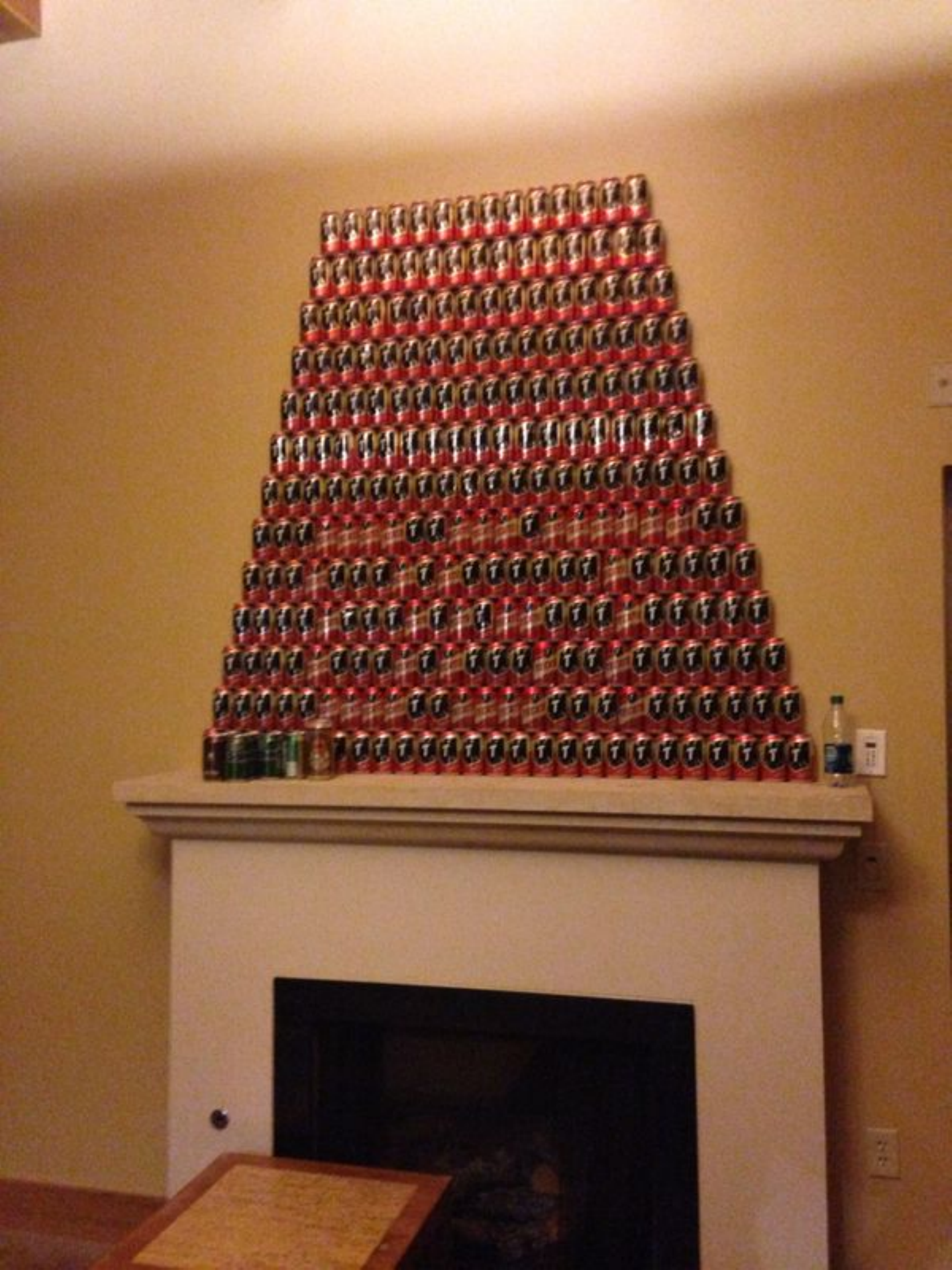}
\caption{\textbf{Please drink responsibly.
}}
\end{figure*}

\end{document}